# Topological Surface States Protected From Backscattering by Chiral Spin Texture

Pedram Roushan[1], Jungpil Seo[1], Colin V. Parker[1], Y. S. Hor[2], D. Hsieh[1], Dong Qian[1], Anthony Richardella[1], M. Z. Hasan[1], R. J. Cava[2] & Ali Yazdani[1]

[1]*Joseph Henry Laboratories & Department of Physics, Princeton University, Princeton, NJ 08544*

[2]*Department of Chemistry, Princeton University, Princeton, NJ 08544.*

**Topological insulators are a new class of insulators in which a bulk gap for electronic excitations is generated by strong spin-orbit coupling[1-5]. These novel materials are distinguished from ordinary insulators by the presence of gapless metallic boundary states, akin to the chiral edge modes in quantum Hall systems, but with unconventional spin textures. Recently, experiments and theoretical efforts have provided strong evidence for both two- and three-dimensional topological insulators and their novel edge and surface states in semiconductor quantum well structures[6-8] and several Bi-based compounds[9-13]. A key characteristic of these spin-textured boundary states is their insensitivity to spin-independent scattering, which protects them from backscattering and localization. These chiral states are potentially useful for spin-based electronics, in which long spin coherence is critical[14], and also for quantum computing applications, where topological protection can enable fault-tolerant information processing[15,16]. Here we use a scanning tunneling microscope (STM) to visualize the gapless surface states of the three-dimensional topological insulator $Bi_{1-x}Sb_x$ and to examine their scattering behavior from disorder caused by random alloying in this compound. Combining STM and angle-resolved photoemission spectroscopy, we show that despite strong atomic scale disorder, backscattering between states of opposite**

**momentum and opposite spin is absent. Our observation of spin-selective scattering demonstrates that the chiral nature of these states protects the spin of the carriers; they therefore have the potential to be used for coherent spin transport in spintronic devices.**

Angle-resolved photoemission spectroscopy (ARPES) experiments on the (111) surface of $Bi_{1-x}Sb_x$ crystals have been used to identify surface states within the bulk band gap of these compounds[9,10]. The shape of the Fermi surface for these states shows an odd number of band crossings between time-reversal invariant momentum points in the first Brillion zone (FBZ) at the Fermi energy, which confirms the identification of $Bi_{1-x}Sb_x$ as a strong topological insulator for $x > 7\%$. The odd number of crossings protects the surface states from being gapped regardless of the position of the chemical potential or the influence of non-magnetic perturbations[4]. Furthermore, spin-sensitive experiments have established that these surface states have a chiral spin structure and an associated Berry's phase[10], which makes them distinct from ordinary surface states with strong spin-orbit coupling[17]. All these characteristics suggest that backscattering, or scattering between states of equal and opposite momentum, which results in Anderson localization in typical low-dimensional systems, will not occur for these two-dimensional carriers. Random alloying in $Bi_{1-x}Sb_x$, which is not present in other material families of topological insulators found to date, makes this material system an ideal candidate to examine the impact of disorder on topological surface states. However, to date there have been no experiments that have probed whether these chiral two-dimensional states are indeed protected from spin-independent scattering.

We performed our experiments using a home-built cryogenic STM that operates at 4 K in ultrahigh vacuum. Single crystal samples of $Bi_{0.92}Sb_{0.08}$ were cleaved *in situ* in UHV at room temperature prior to STM experiments at low temperatures. The topographic images of the sample are dominated by long wavelength (~20Å) modulations in the local density of states (Fig. 1a). However, atomic corrugation can also be observed in the topography (inset Fig. 1a). Spectroscopic measurements show a general suppression of the density of states near the Fermi level, with spectra appearing for the most part homogenously across the sample surface (Fig. 1b). ARPES measurements[9,10] and recent band structure calculations[13] suggest that within ±30meV of the Fermi level, where there is a bulk gap, tunneling should be dominated by the surface states. While tunneling spectroscopy measurements do not distinguish between bulk and surface states, energy-resolved spectroscopic maps shown in Fig.'s 2a, b and c display modulations that are the result of scattering of the surface electronic states. As expected for the scattering and interference of surface states, the observed patterns are not commensurate with the underlying atomic structure. While we do not have direct information on the identity of the scattering defects, the random distribution of substituted Sb atoms is a likely candidate.

Energy-resolved Fourier transform scanning tunneling spectroscopy (FT-STS) can be used to reveal the wavelengths of the modulations in the local density of states and to obtain detailed information on the nature of scattering processes for the surface state electrons[18]. Previous studies on noble metal surface states[18,19] and high-$T_C$ superconductors[20] have established the link between modulation in the conductance at a wavevector $\vec{q}$ and elastic scattering of quasi-particles and their interference between different momentum states ($\vec{k_1}$ and $\vec{k_2}$, where $\vec{q} = \vec{k_1} + \vec{k_2}$) at the same energy. The FT-STS maps shown as insets in Fig.'s 1a,b and c for $Bi_{0.92}Sb_{0.08}$ display rich qausi-particle interference (QPI) patterns, which have the six-fold rotational symmetry of the underlying lattice, and evolve as a function of energy. These patterns display the allowed wavevectors $\vec{q}$ and the relative intensities for the various scattering processes experienced by the surface state electrons.

Within a simple model of QPI, the interference wavevectors connect regions of high density of states on contours of constant energy (or the Fermi surface at the chemical potential). Therefore the QPI patterns should correspond to a joint density of states (JDOS) for the surface state electrons that can be independently determined from ARPES measurements[20-23]. Fig. 2d and e show contours of constant energy (CCE) in the FBZ, as measured with ARPES at two energies on $Bi_{0.92}Sb_{0.08}$ crystals (following procedures described in Ref 9). The CCE consist of an electron pocket centered on the $\overline{\Gamma}$ point, hole pockets half way to the $\overline{M}$ point, and two electron pockets that occur very close to the $\overline{M}$ point[9,10]. From these measurements, we can determine the JDOS as a function of the momentum difference between initial and final scattering states, $\vec{q}$, using

$$JDOS(\vec{q}) = \int I(\vec{k})I(\vec{k}+\vec{q})d^2\vec{k} \text{ , where } I(\vec{k}) \text{ is the ARPES intensity that is proportional}$$

to the surface states' density of states at a specific two-dimensional momentum $\vec{k}$. Fig. 3a and b shows the results of computation of the JDOS from ARPES data for two different energies. Contrasting these figures to the corresponding QPI data in Fig. 3b and 3e, we find a significant suppression of the scattering intensity along the directions equivalent to $\overline{\Gamma}$ - $\overline{M}$ in the FBZ. Backscattering between various electron and hole pockets around the $\overline{\Gamma}$ point should give rise to a continuous range of scattering wavevectors along the $\overline{\Gamma}$ - $\overline{M}$ direction, a behavior not observed in the data (see also the expanded view of the JDOS and QPI in Fig. 4a). This discrepancy suggests the potential importance of the surface states' spin texture and the possibility that spin rules are limiting the backscattering for these chiral electronic states.

To include spin effects, we use the results of spin-resolved ARPES studies and assign a chiral spin texture to the electron and hole pockets as shown in Fig. 2f. ARPES studies have resolved the spin structure only for the central electron pocket and the hole pockets near the $\overline{\Gamma}$ point in the FBZ; however, we assign a similar chiral structure for states near $\overline{M}$. This assignment is consistent with the presence of a $\pi$ Berry's phase that distinguishes the spin topology of the $Bi_{0.92}Sb_{0.08}$ surfaces states from that of surface state bands that are simply split by spin-orbit coupling. In the latter case, the spin-polarized surface bands come in pairs, while for topological surface states there should be an odd number of spin-polarized states between two time reverse equivalent points in the band structure.[1,2,11,13]

To understand scattering and interference for these spin-polarized states, we determine the spin-dependent scattering probability, $SSP(\vec{q}) = \int I(\vec{k}) T(\vec{q},\vec{k}) I(\vec{k}+\vec{q}) d^2\vec{k}$, which in similar fashion to the JDOS uses the ARPES-measured density of states, $I(\vec{k})$, but also includes a spin-dependent scattering matrix element $T(\vec{q},\vec{k})$. This matrix element describes the scattering probability as a function of momentum transfer and spins of states that are connected by the scattering process. Shown in Fig. 3c and f are the calculated $SSP(\vec{q})$ from ARPES data at two different energies using a matrix element of the form $T(\vec{q},\vec{k}) = \left| \left\langle \vec{S}(\vec{k}) \middle| \vec{S}(\vec{k}+\vec{q}) \right\rangle \right|^2$. This simple form of spin-selective scattering reduces scattering between states with nonaligned spins and completely suppresses scattering between states with opposite spin orientations. Comparison of the SSP patterns to the QPI measurements in Fig. 3 shows that including spin effects leads to remarkably good agreement between the scattering wavevectors measured by STM and those expected from the shape of the surface CCE as measured by ARPES (features in the FT-STS and SSP at different wavevectors are categorized and given labels in Fig. 3g). A quantitative comparison between the QPI from the STM data and JDOS and SSP from ARPES data can be made by computing the cross correlation between the various patterns. Focusing on the high-symmetry direction, which is shown in Fig. 4a, we find that the QPI (excluding the central q=0 section, which is dominated by the disorder) is 95% correlated with the SSP in the same region. The cross-correlation is found to be 83% between the QPI and JDOS. Therefore, the proposed form of the spin-dependent scattering matrix element is the critical component for understanding the suppression of scattering along the high symmetry directions in the data.

The proposed scattering matrix elements $T(\vec{q}, \vec{k})$ and associated spin-scattering rules are further confirmed by a more comprehensive analysis of the QPI patterns. An example of such an analysis is shown in Fig. 4, in which we associate features along the high symmetry direction in the QPI and SSP with specific scattering wavevectors $\vec{q}$ that connect various regions of the CCE. The observed wavevectors in QPI and SSP obey spin rules imposed by $T(\vec{q}, \vec{k})$, as illustrated schematically in Fig. 4b. We also depict in Fig. 4b examples of scattering processes that, while allowed by the band structure and observed in JDOS, violate the spin scattering rules and are not seen in QPI data in Fig. 4a. A comprehensive analysis of all the features in the QPI data (included in the supplementary section) demonstrates that allowed set of scattering wavevectors $\vec{q}_1$ through $\vec{q}_8$ (Fig. 3g) exclude those that connect states with opposite spin. Finally, in Fig. 5, we show dispersion as a function of energy for some of the wavevectors $\vec{q}$ in the QPI and compare their energy dispersion to that expected from ARPES results in the SSP. Remarkably, all the features of the complex QPI patterns and their energy dependence can be understood in detail by the allowed scattering wavevectors based on the band structure of the topological surface states and the spin scattering rule. This agreement provides a precise demonstration that scattering of electrons over thousands of Angstroms, which underlies the QPI maps, strictly obeys the spin scattering rules and associated suppression of backscattering.

Other surface states with strong spin-orbit interaction may be expected to show evidence for spin-selective scattering; however, since spin states come in pairs, the QPI patterns can rarely probe these rules[24]. In some situations there is evidence of such rules[25,26], but the precision with which scattering of surface states for $Bi_{0.92}Sb_{0.08}$ can be understood using spin-selective scattering is unprecedented. Unusual scattering of chiral electronic states is also seen in monolayer graphene, where the underlying two-atom basis leads to a pseudospin index for quasi-particles and results in suppression of intravalley scattering[27,28]. The key difference expected for surface states of a topological insulator is the degree to which they can tolerate disorder. This aspect is clearly demonstrated here for surface states of $Bi_{0.92}Sb_{0.08}$, where strong alloying causes scattering for the surface state electrons yet the spin-selection scattering rules are strictly obeyed over length scales much longer than that set by the atomic scale disorder. Future experiments with magnetic scattering centers can further probe the spin scattering rules for topological surface states and may provide the setting for the manipulation of these spin-polarized states in device applications.

**Methods Summary**

The $Bi_{0.92}Sb_{0.08}$ single crystals were grown by melting stoichiometric mixtures of elemental Bi(99.999%) and Sb(99.999%) in 4mm inner diameter quartz tubes. The samples were cooled over a period of two days, from 650 to 260 ˚C, and then annealed for a week at 260 ˚C. The samples were cleaved *in situ* in our home-built cryogenic STM that operates at 4 K in ultra-high-vacuum. The STM topographies were obtained in constant-current mode, and dI/dV spectroscopy was measured by a standard lock-in technique with f=757Hz, and an AC modulation of 3mV added to the bias voltage. The spatial resolution during the dI/dV mapping was about 2Å, which provided the capability to resolve k-vectors up to twice the first Brillion zone in momentum space. The spin-resolved ARPES measurements were performed at the SIS beam line at the Swiss Light Source using the COPHEE spectrometer with a single 40 kV classical Mott detector and photon energies of 20 and 22 eV. The typical energy resolution was 80mV, and momentum resolution was 3% of the surface BZ. The JDOS formula provides a map of all scatterings by calculating the self-convolution of a given CCE. While JDOS disregards the spin texture of a CCE, the SSP formula considers a weight factor for each possible scattering proportional to the square of the projection of the initial spin state onto the final state.

We gratefully acknowledge K. K. Gomes and A.N. Pasupathy for important suggestions on experimental procedure and initial analysis. This work was supported by grants from ONR, ARO, NSF-DMR, and NSF-MRSEC programme through Princeton Center for Complex Materials. P.R. acknowledges support of NSF graduate fellowship.

**Author Contributions** Y. S. Hor and R. J. Cava carried out the growth of the single crystals and characterized them. D. Hsieh, D. Qian and M.Z. Hasan performed the ARPES studies of the samples. STM measurements as well as data analysis was done by Pedram Roushan, Jungpil Seo, Colin V. Parker, Anthony Richardella and  Ali Yazdani.

Correspondence and requests for materials should be addressed to A.Y. (yazdani@princeton.edu).

**Figure 1. STM topography, and dI/dV spectroscopy of the $Bi_{0.92}Sb_{0.08}$ (111) surface. a,** STM topograph (+50meV, 100pA) of the $Bi_{0.92}Sb_{0.08}$ (111) surface over an 800 Å by 800 Å area. The inset shows an area of 80 Å by 80 Å that displays the underlying atomic lattice (+200mV, 15pA). **b,** Spatial variation of the differential conductance (dI/dV) measurements across a line of length 250Å.

A typical differential conductance measurement over larger energy ranges is shown in the inset.

**Figure 2. dI/dV maps, QPI patterns, and ARPES measurements on Bi$_{0.92}$Sb$_{0.08}$ (111) surface. a, b,** and **c,** Spatially resolved conductance maps of the Bi$_{0.92}$Sb$_{0.08}$ (111) surface obtained at -20 mV, 0 mV, and +20 mV (1000 Å x 1300 Å). In the upper right corner of each map the Fourier transform of the dI/dV maps are presented. The hexagons have the same size as the FBZ. The Fourier transforms have been symmetrized in consideration of the three-fold rotation symmetry of the (111) surface. **d** and **e,** ARPES intensity map of the surface state at -20mV and at the Fermi level, respectively. **f,** The spin textures from ARPES measurements are shown with arrows, and high symmetry points are marked ($\overline{\Gamma}$ and 3 $\overline{M}$ ).

**Figure 3. Construciton of joint density of states (JDOS) and spin scattering probability (SSP) from ARPES data and their comparison with FT-STS.**

**a,** the JDOS and SSP calculated at E$_F$, from ARPES data presented in Fig. **2e.** **b,** the FT-STS at E$_F$. **c,** the SSP calculated at E$_F$. **d,** the JDOS calculated at -20mV, from ARPES data presented in Fig. **2d. e,** the FT-STS at -20mV. **f,** the SSP calculated at -20mV. **g,** the schematization of the features associated with scattering wavevectors $\vec{q}_1$ to $\vec{q}_8$ in the FT-STS data.

**Figure 4. Comparison of the various parts of the QPI patterns along the $\overline{\Gamma}$ - $\overline{M}$ direction at Fermi level. a,** close up view of the QPI pattern from JDOS, FT-STS, and SSP at Fermi level, along the $\overline{\Gamma}$ - $\overline{M}$ direction. The last row shows the schematic representation of q$_2$,q$_4$, and q$_8$ , which correspond to scatterings shown in **b**. Two high intensity points which are only seen in JDOS are labeled as A and B. **b,** the Fermi surface along the $\overline{\Gamma}$ - $\overline{M}$ direction, with spin orientation

of the quasiparticles shown with arrows. The horizontal color-coded arrows show the sources of the scatterings seen in the STM data. Note that all highlighted spins have the same orientation. The top row depicts the scatterings which involve opposite spins and are presented in JDOS, but absent in FT-STS, and SSP.

**Figure 5. Dispersion of various peaks from FT-STS and ARPES. a**, the intensity of the FT-STS maps along the $\overline{\Gamma}$ - $\overline{M}$ direction for various energies. The two peak positions correspond to $q_1$ and $q_2$, which become larger with increasing in energy. Each curve is shifted by 0.6 pS for clarity. **b,** dispersion of the position of $q_1$, $q_2$, and $q_3$ from ARPES (open symbols) and STM (solid symbols). The data were obtained from fitting the peak in the intensity of the QPI patterns measured in STM, and calculated from the ARPES CCE. Each STM data point is the averaged value of six independent measurements, and the error bar represents one standard deviation. The systematic error was negligible.

*Methods*

***Crystal growth*** The $Bi_{0.92}Sb_{0.08}$ single crystals were grown by melting stoichiometric mixtures of elemental Bi(99.999%) and Sb(99.999%) in 4mm inner diameter quartz tubes from a stoichiometric mixture of high purity elements. The samples were cooled over a period of two days, from 650 to 260 ˚C, and then annealed for a week at 260 ˚C. The obtained single crystals were confirmed to be single phase and identified as having the rhombohedral A7 crystal structure by X-ray power diffraction using a Bruker D8 diffractometer with Cu Kα radiation and a graphite diffracted beam monochromator.

***STM measurement*** We performed our experiments using a home-built cryogenic STM that operates at 4 K in ultrahigh vacuum. With our STM, we have examined several crystals of $Bi_{0.92}Sb_{0.08}$, grown under the same conditions, and we have not noticed any sample dependence for any of the results we are presenting in this article. The typical size of the crystals we used was 1 mm×1 mm×0.3 mm. Samples were cleaved *in situ* at room temperature in UHV prior to STM experiments at low temperatures. The weak bonding between bilayers in this crystal makes the (111) surface the natural cleavage plane. A mechanically sharpened Pt–Ir alloy wire was used as an STM tip, and the quality of the tip apex was examined by scanning an atomically clean Ag(111) surface.

The STM topographies were obtained in constant-current mode, and dI/dV spectroscopy was measured by a standard lock-in technique with f=757Hz, an ac modulation of 3mV added to the bias voltage, and the feed-back loop disabled during the measurement. The spatial resolution during the dI/dV mapping was about 2Å, which provides the capability to resolve k-vectors up to twice the first Brillion zone in momentum space. The real space resolution also guarantees inclusion of atomic peaks in FT-STS maps, providing the most accurate calibration. In addition, the deviation of these atomic peaks from a perfect hexagon can be used as a measure of the thermal drift, which for the results presented was negligible, allowing us to symmetrize the FT-STS results without smearing out features or creating artificial ones.

***ARPES measurements*** High resolution ARPES and spin-resolved ARPES have been measured in different labs. High-resolution ARPES data have been taken at beamlines 12.0.1 and 10.0.1 of the Advanced Light Source at the Lawrence Berkeley National Laboratory, as well as at the PGM beamline of the Synchrotron Radiation Center in Wisconsin, with photon energies ranging from 17 eV to 55 eV and energy resolutions ranging from 9 meV to 40 meV, and momentum resolution better than 1.5% of the surface Brillouin zone (BZ). Spin-integrated ARPES measurements were performed with 14 to 30 eV photons on beam line 5-4 at the Stanford Synchrotron Radiation Laboratory, and with 28 to 32 eV photons on beam line 12 at the Advanced Light Source, both endstations being equipped with a Scienta hemispherical electron analyzer (see VG Scienta manufacturer website for instrument specifications).

Spin-resolved ARPES measurements were performed at the SIS beam line at the Swiss Light Source using the COPHEE spectrometer with a single 40 kV classical Mott detector and photon energies of 20 and 22 eV. The typical energy and momentum resolution was 15 meV and 1.5% of the surface BZ respectively at beam line 5-4, 9 meV and 1% of the surface BZ respectively at beam line 12, and 80 meV and 3% of the surface BZ respectively at SIS using pass energy of 3 eV.

***JDOS and SSP calculation***   Two plausible assumptions were considered for scattering between quasiparticles, spin independent (JDOS) and spin dependent scattering (SSP). The JDOS formula provides a map of all scatterings by calculating the self-convolution of a given CCE, and hence disregards the spin texture of a CCE. The SSP formula considers a weight factor for each possible scattering, proportional to the square of the projection of the initial spin state onto the final state. Therefore, vectors connecting quasiparticles of opposite spins do not contribute to the SSP mapping. The mean subtracted correlation of the FT-STS to the JDOS and SSP was calculated in a rectangular region starting $0.08\text{Å}^{-1}$ away from the $\overline{\Gamma}$ point and ending at $\overline{M}$, and of width $0.22\text{Å}^{-1}$. The calculation of SSP using various subsets of the pockets of the Fermi surface enabled us to identify scattering processes which give rise to the various q's (*see supplementary Fig. 1S*).  By using a Gaussian fit to the peak of the intensity, we were able to follow the energy evolution of several q's in the FT-STS maps, (*see fig. 5 of the main article*). A similar fitting procedure was applied to the ARPES intensity maps at fixed energies.

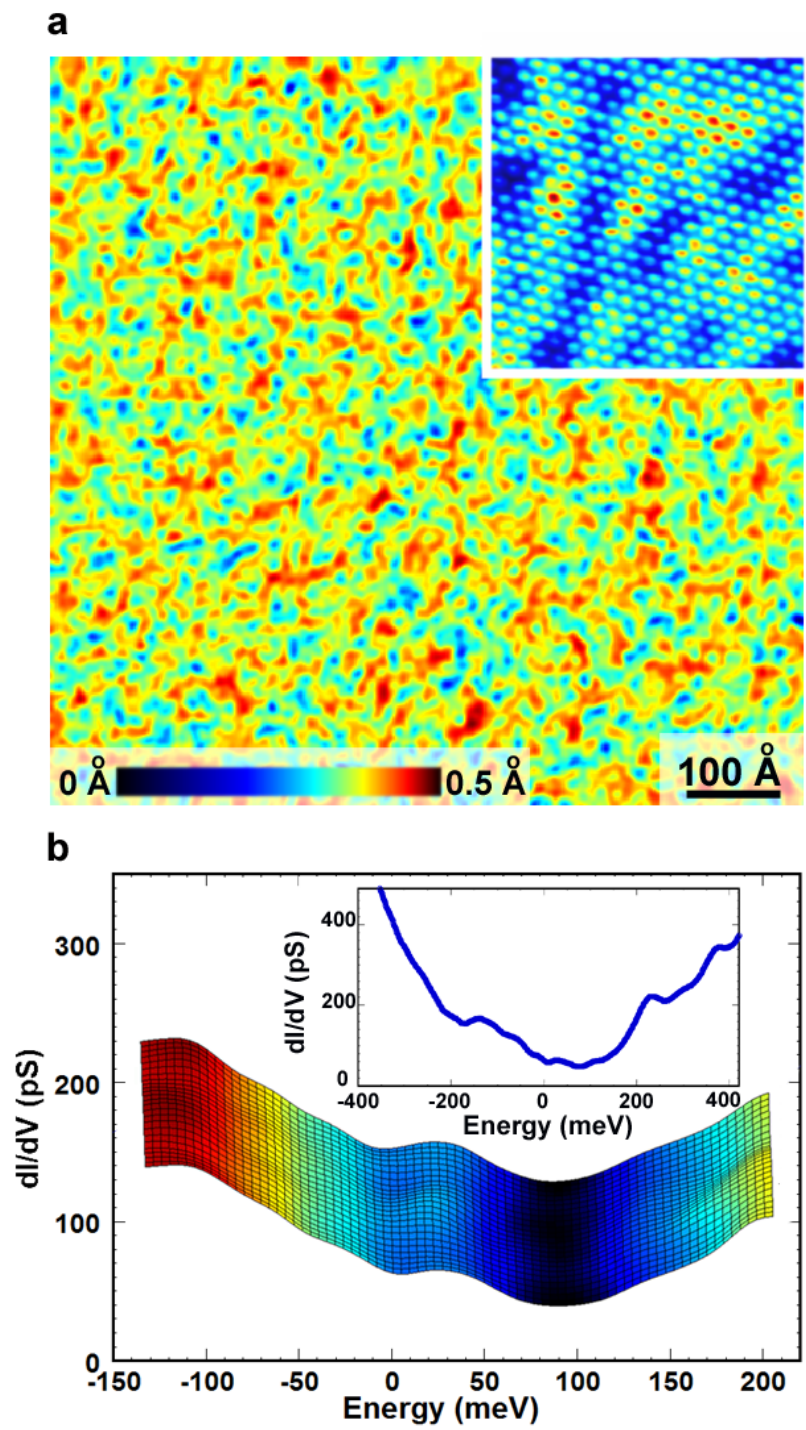

Figure 1

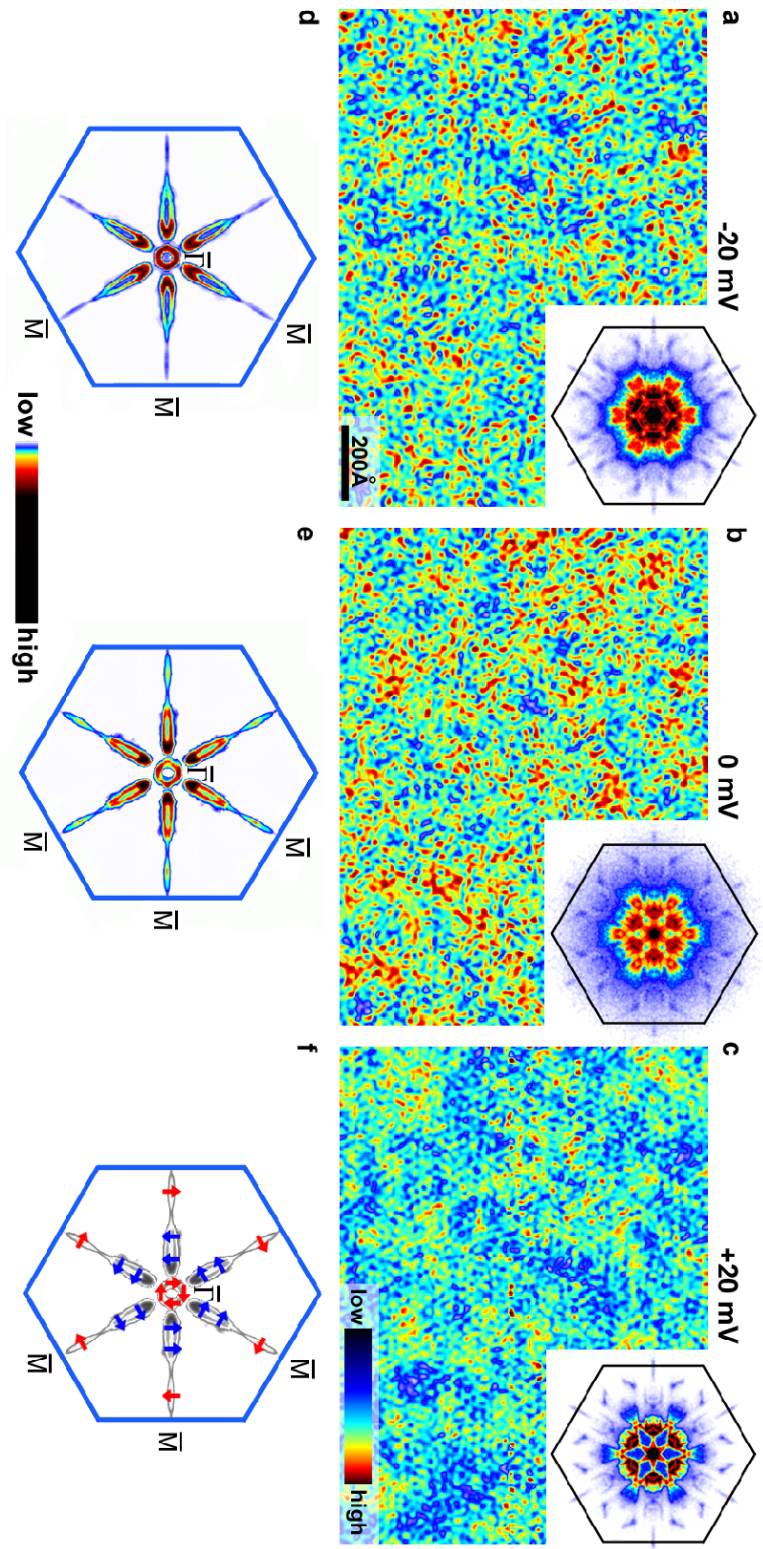

Figure 2

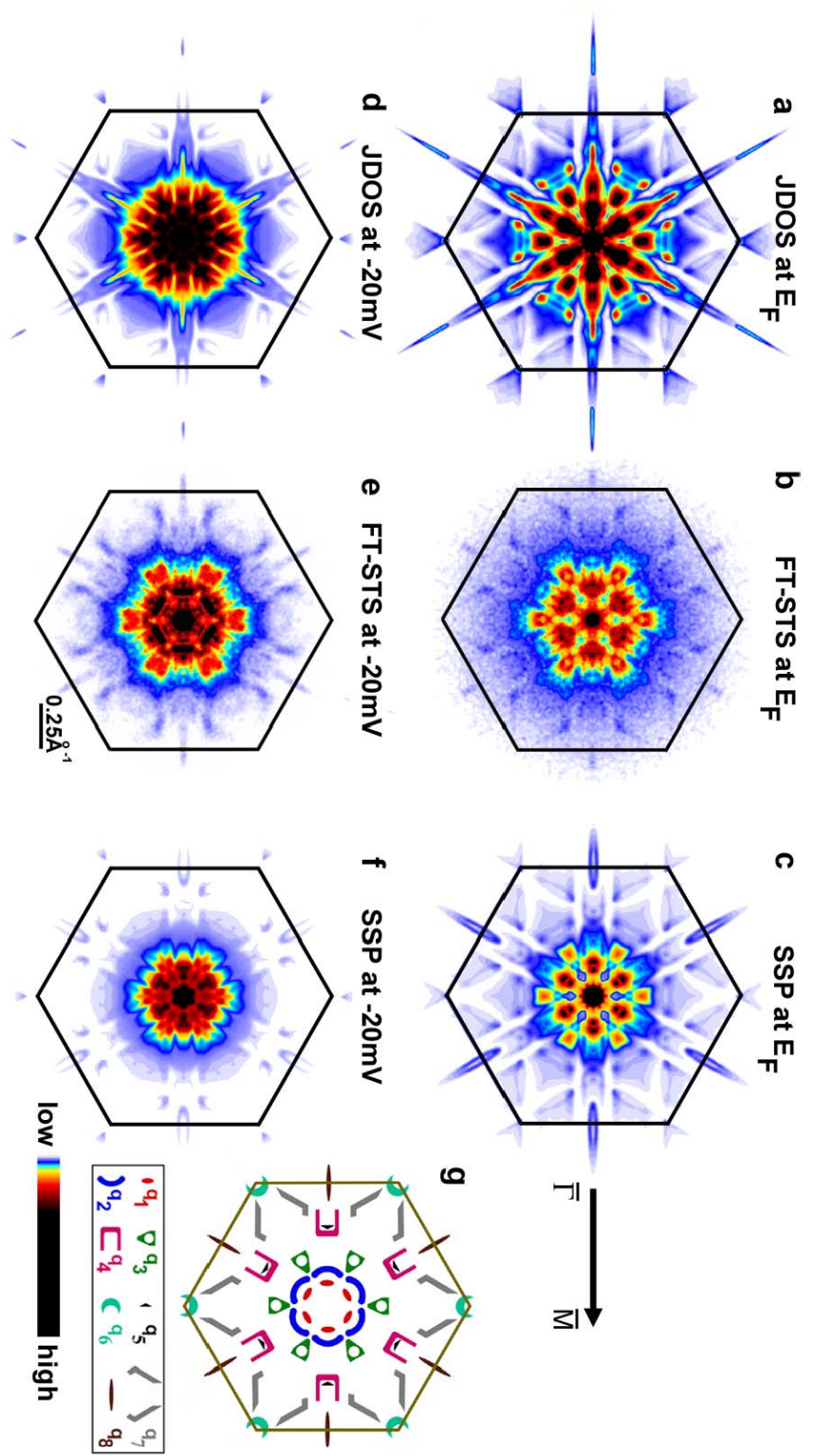

Figure 3

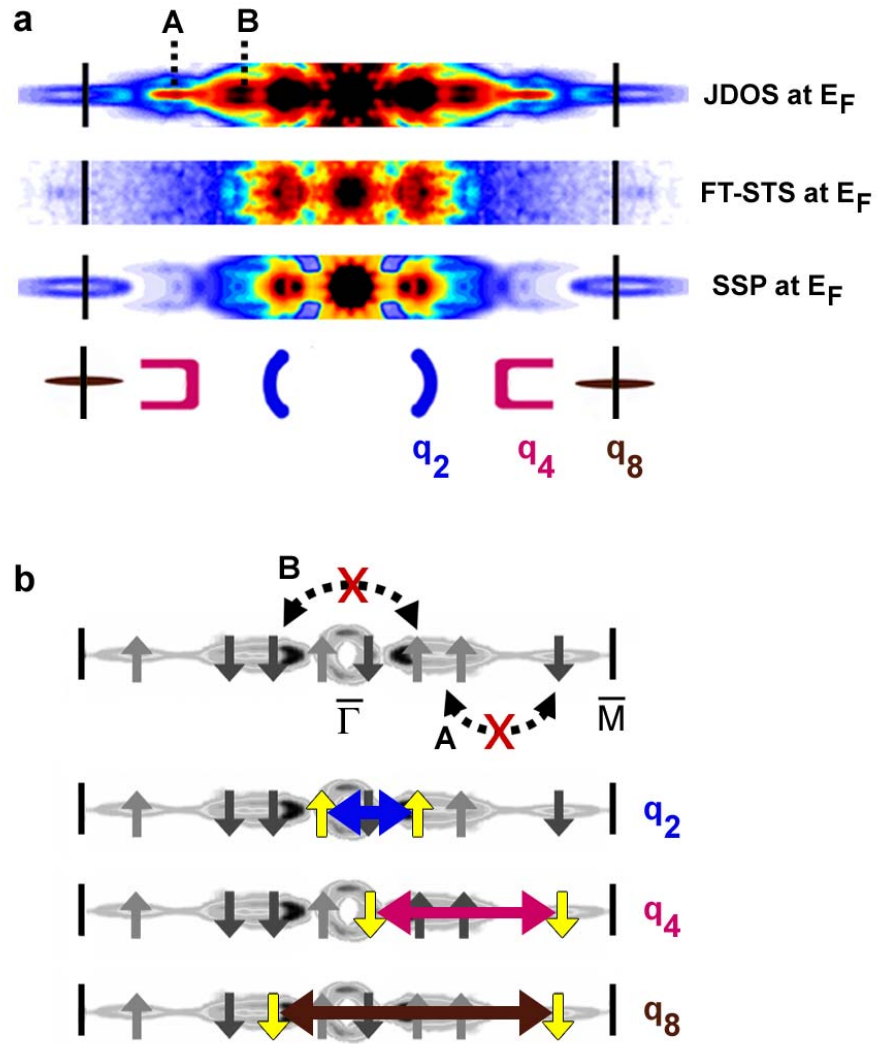

**Figure 4**

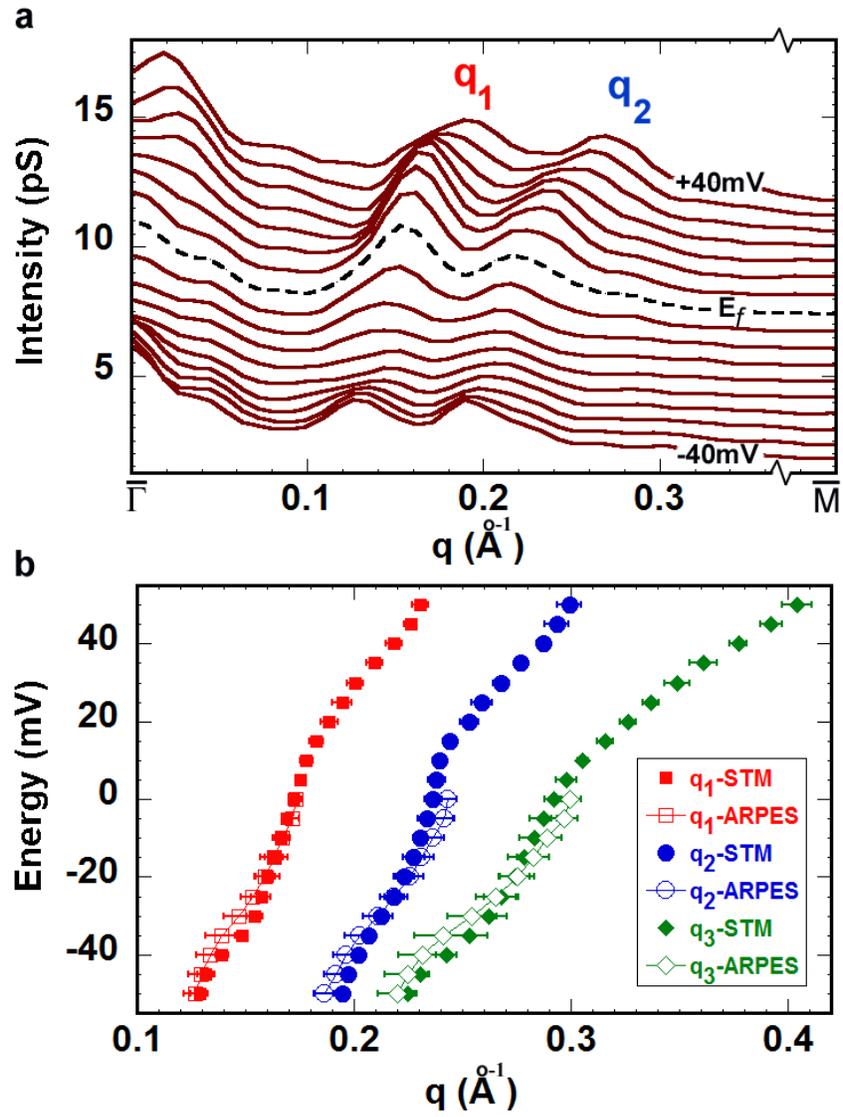

**Supplementary figure**

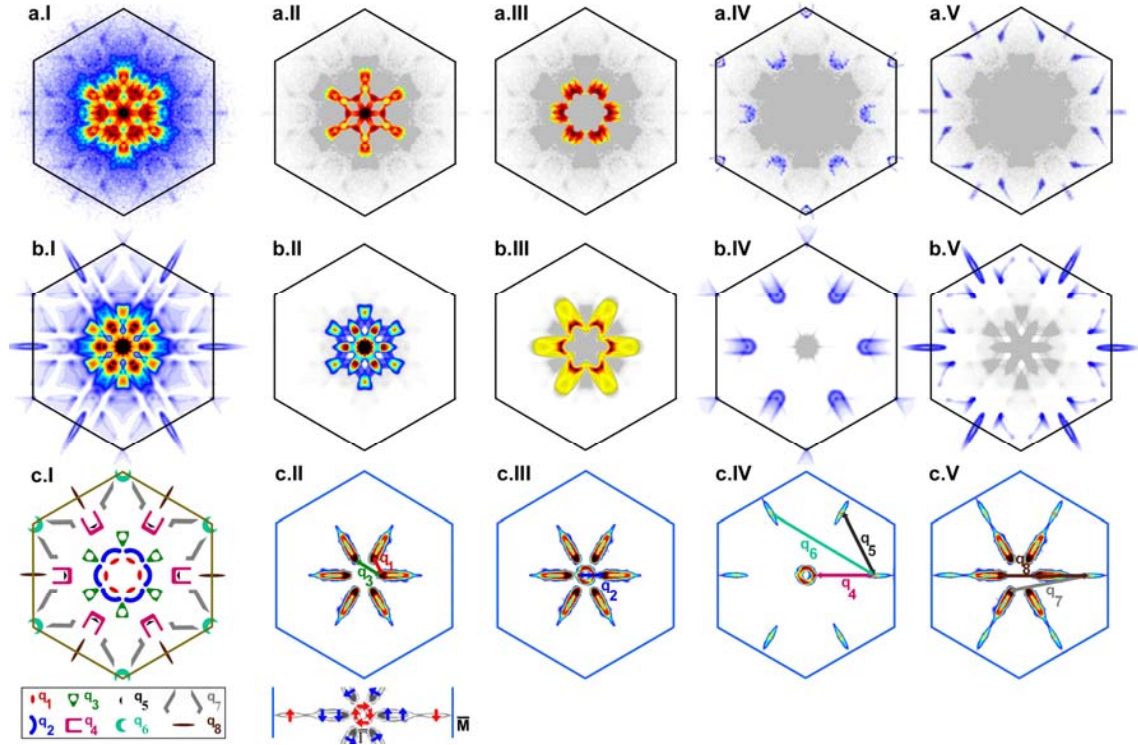

**Figure S1. Decomposition of the various parts of the QPI pattern at the Fermi level. a.I,** the Fermi level QPI measured with STM. The individual features are highlighted in **a.II** to **a.V**. **b.I,** the SSP calculation of the QPI pattern from ARPES intensity maps at Fermi level, and its decomposition into various constituent parts is shown in **b.II** to **b.V**. **c.I,** the schematization of the various features seen in the FT-STS data. **c.II** to **c.V,** various parts of the Fermi contours measured by ARPES, with arrows showing the sources of the scatterings seen in STM data. Columns **II** to **V** have the following order: in **c**, we show a specific part of the Fermi surface, and in **b** the SSP based of that part is presented. In **a**, the corresponding parts of the pattern which are visible in STM data are in color and the rest are shown in gray. In the legend, the ARPES intensity map and its spin texture at the Fermi level is shown.